\documentclass{llncs}
\usepackage[T1]{fontenc}
\usepackage[colorlinks=true, linkcolor=blue, citecolor=blue, urlcolor=blue]{hyperref}

\usepackage{fancyhdr}

\usepackage{multirow}
\usepackage{hyperref}
\usepackage{enumitem}
\usepackage{graphicx}
\usepackage{subcaption}
\usepackage{tcolorbox}
\usepackage{graphicx}
\usepackage{array}
\usepackage{longtable}
\usepackage{lscape}
\usepackage{adjustbox}
\usepackage{makecell}
\usepackage{orcidlink}

\usepackage{xcolor}

\graphicspath{ {./fig/}  } 

\begin{document}
\title{What Makes Teamwork Work? A Multimodal Case Study on Emotions and Diagnostic Expertise in an Intelligent Tutoring System}

%
%
\author{}
\institute{}


\author{
Xiaoshan Huang\orcidlink{0000-0002-2853-7219} \and
Haolun Wu\orcidlink{0000-0001-6255-1535} \and
Xue Liu\orcidlink{0000-0001-5252-3442} \and
Susanne P. Lajoie\orcidlink{0000-0003-2814-3962}
}
\institute{
McGill University, Montreal, Quebec, Canada\\
\email{\{xiaoshan.huang, haolun.wu\}@mail.mcgill.ca}\\
\email{\{xue.liu, susanne.lajoie\}@mcgill.ca}
}

%
\maketitle              
\begin{abstract}
Teamwork is pivotal in medical teamwork when professionals with diverse skills and emotional states collaborate to make critical decisions. This case study examines the interplay between emotions and professional skills in group decision-making during collaborative medical diagnosis within an Intelligent Tutoring System (ITS). By comparing verbal and physiological data between high-performing and low-performing teams of medical professionals working on a patient case within the ITS, alongside individuals’ retrospective collaboration experiences, we employ multimodal data analysis to identify patterns in team emotional climate and their impact on diagnostic efficiency. Specifically, we investigate how emotion-driven dialogue and professional expertise influence both the information-seeking process and the final diagnostic decisions. Grounded in the socially shared regulation of learning framework and utilizing sentiment analysis, we found that social-motivational interactions are key drivers of a positive team emotional climate. Furthermore, through content analysis of dialogue and physiological signals to pinpoint emotional fluctuations, we identify episodes where knowledge exchange and skill acquisition are most likely to occur. Our findings offer valuable insights into optimizing group collaboration in medical contexts by harmonizing emotional dynamics with adaptive strategies for effective decision-making, ultimately enhancing diagnostic accuracy and teamwork effectiveness.
\end{abstract}
\section{Introduction}
\vspace{-2mm}
Effective teamwork allows groups to achieve more than what individuals working in isolation can accomplish. However, working together does not guarantee optimal outcomes~\cite{salas2005there}. Collaborative dynamics is essential in team success, especially when emotions are contagious and shared through social interactions~\cite{andersen1996principles}.

Educational technologies support learners' knowledge acquisition~\cite{jerinic2000friendly}, where socio-emotional interactions are crucial to their knowledge co-construction~\cite{huang2023social,huang2024examining}. Challenges remain in understanding teamwork within technology-rich environments, where interactions involve not only human-to-human dynamics but also human-computer interactions. These interactions introduce complexity, and multimodal data analysis~\cite{huang2024scoping}, which captures effects through various channels such as speech (e.g., emotional tones)~\cite{dehbozorgi2021aspect} and physiological responses (e.g., heart rate variabilities)~\cite{donker2023using,appelhans2006heart}, offers avenues to understand these nuanced dynamics~\cite{zhao2023analysing}.
\vspace{-2mm}
\section{Background}
\vspace{-2mm}
\subsection{The Role of Affective States in Teamwork} 
\vspace{-1mm}
Teamwork operates on both cognitive and affective dimensions, with emotions playing a pivotal, yet less-understood role in shaping team dynamics and overall performance. Positive affective states, such as optimism or curiosity, have been shown to enhance team engagement, foster collaboration, and improve problem-solving efficacy~\cite{haas2016secrets}. On the other hand, negative emotions can disrupt communication, hinder decision-making, and ultimately reduce team effectiveness~\cite{polo2022affording}. Understanding the intricate relationship between emotions and teamwork is critical, particularly in complex tasks within technology-rich environments.

Team emotional climate, the shared affective states within a team have impacts on collaborative experiences~\cite{somech2013translating,kelly2001mood} and team effectiveness~\cite{farh2012emotional}. Emotional tone, expressed through vocal intonation or written communication~\cite{syrjamaki2023emotionally}, reflects individuals' emotional states, which in turn influences cognitive processes~\cite{chang2023emotional}.  It can significantly affect team performance, highlighting the importance of considering affective factors alongside cognitive ones~\cite{imai2010emotions}.  

\vspace{-2mm}
\subsection{Interactions and Team Emotional Climate in Teamwork}  
\vspace{-1mm}
Grounded from the Socially Shared Regulation of Learning framework (SSRL~\cite{panadero2015socially}) we argue that team effectiveness hinges on shared regulation of emotions, motivation, cognition and metacognition~\cite{jarvela2016socially}. SSRL becomes crucial when meaningful interactions are elicited among team~\cite{jarvela2023predicting,nguyen2023examining}. 

These interactions are closely associated with active participation and overall success in teamwork~\cite{isohatala2017socially}. Metacognitive interaction facilitates justification of behavior to one and others~\cite{frith2012role}, enhancing team cohesion and performance. Socio-cognitive interactions, such as monitoring and reflection, are also strongly linked to team success. Emotional and motivational interactions, while closely related to socio-cognitive interactions, are more challenging to distinguish and measure~\cite{naykki2017facilitating}. The complexity of human-computer interactions adds another layer to these interactions, affecting learner engagement dynamically~\cite{chen2024identifying,huang2023relative}. Emotional contagion, which shapes team emotional climate, plays a vital role in influencing team efficiency. Analyzing emotional sharing and regulation thus offers a key lens for optimizing teamwork~\cite{perry2013tracing}. 

In this case study, we investigated two research questions (RQs) are raised within an Intelligent Tutoring System (ITS) context:  
\textit{\textbf{1.} How do interaction types relate to a team's emotional climate?} 
\textit{\textbf{2.} What factors trigger emotional changes during team decision-making?}

\vspace{-2mm}
\section{Methodology}
\vspace{-2mm}
\subsection{Participants and Learning Environment}
\vspace{-1mm}
This case study consisted of 4 dyads of female medical professionals (Mean$_{age}$ = $28.13$) enrolled in a residency program at a large North American university. The study was conducted after receiving approval from the ethics board of the institute where participants were involved. BioWorld~\cite{lajoie2021student,huang2023relative}, an intelligent tutoring system designed to assist medical students in learning through problem-solving tasks, was used as the media and learning platform. Each dyad was co-located to share the same screen. The BioWorld interface includes four primary functions: \textit{Patient Case}, \textit{Tests}, \textit{Library}, and \textit{Hypotheses}, accessible via a left-side navigation bar where learners can freely navigate between these functions. 

\vspace{-2mm}
\subsection{Measurements}
\vspace{-1mm}
Verbal data were collected through transcribed dialogues from each team. Team emotional climate was quantified using compound sentiment scores derived from sentiment analysis, while various interaction types were identified within the dialogue transcriptions.

Physiological data were obtained during the diagnostic task. Participants were fitted with the Empatica E4 bracelet, a non-intrusive device, on their non-dominant hands. Average heart rate was derived from the raw blood volume pulse (BVP) signal, sampled at a default rate of 1 Hz. Heart rate change points were analyzed to indicate moments of heightened emotional intensity or stress, reflecting shifts in
the team’s emotional climate~\cite{pijeira2019sympathetic}. We utilized the Bayesian Estimator of Abrupt change, Seasonal change, and Trend (BEAST) algorithm to probabilistically identify periods of substantial heart rate variation~\cite{zhao2019detecting}. This approach helps to pinpoint moments of physiological arousal changes linked to teamwork experience. Analysis was conducted using the RBeast\footnote{\href{https://pypi.org/project/Rbeast/}{https://pypi.org/project/Rbeast/}}. 

Team effectiveness was measured using objective performance indices including diagnostic efficiency and subjective reflections gathered through semi-structured interviews.

\vspace{-2mm}
\subsection{Analytic Method}
\vspace{-1mm}
We used sentiment analysis to identify team emotional climate~\cite{hutto2014vader}. The compound score serves as an aggregated measure, encapsulating the sentiment polarity of the text, ranging from -1 (extreme negativity) to +1 (extreme positivity).

We adapted a SSRL scheme\footnote{\href{https://github.com/katherine-huang/AIED-2025/blob/main/coding_scheme_SSRL.png}{https://github.com/katherine-huang/AIED-2025/blob/main/coding\_scheme\_SSRL.png}} to code the dialogue. Specifically, we categorized interactions into four types: \textit{Metacognitive}, \textit{Cognitive}, \textit{Emotional}, and \textit{Motivational}. Two researchers coded the dialogues and solved discrepancies among codes. The interrater reliability for each interaction category exceeded 0.7, meeting the criterion of Cohen’s kappa~\cite{mchugh2012interrater}. Content analysis using multimodal data was conducted by linking heart rate change point timestamps to team dialogues, connecting them to specific interactions.

\vspace{-2mm}
\section{Result} 
\vspace{-2mm}
\subsection{RQ1. Interaction and Team Emotional Climate}
\vspace{-1mm}

The dialogue in teams shows an overall neutral to slightly positive emotional climate during the task (Figure 1), with a statistically significant effect of interaction type on emotional tone, \textit{F}(3, 1116) = 4.96, \textit{p} = .002. The compound sentiment of socio-motivational interactions (mean = 0.28), is significantly higher than both cognitive (mean = 0.11, \textit{p} = .02) and metacognitive interactions (mean = 0.16, \textit{p} = .01) based on Tukey's Honestly Significant Difference (HSD) tests.  
\begin{figure}[t]
    \centering
    \begin{subfigure}[b]{0.48\textwidth}
        \centering
        \includegraphics[width=\textwidth]{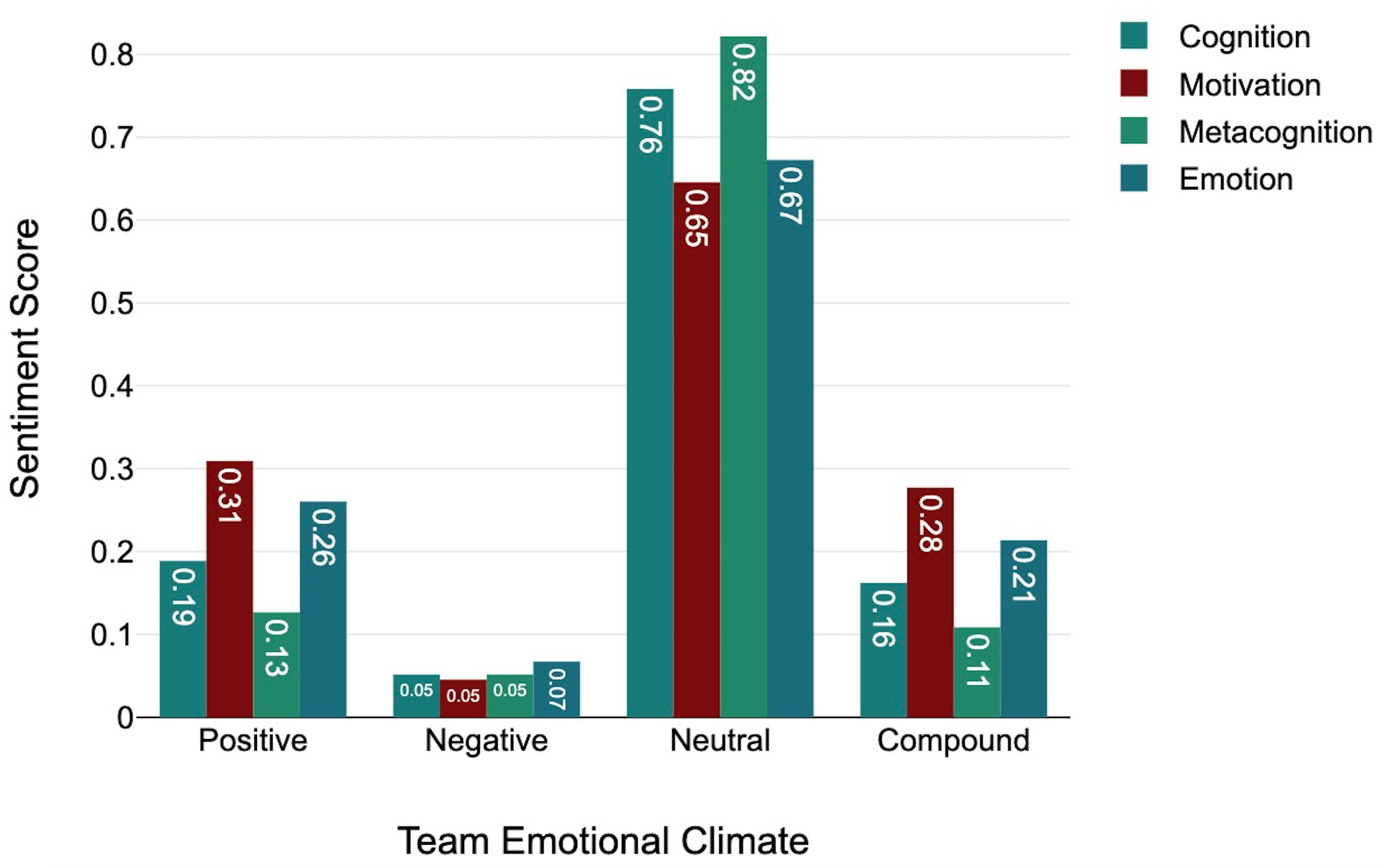} 
        \caption{Interaction by Team Climate}
    \end{subfigure}
    \hfill
    \begin{subfigure}[b]{0.48\textwidth}
        \centering
        \includegraphics[width=\textwidth]{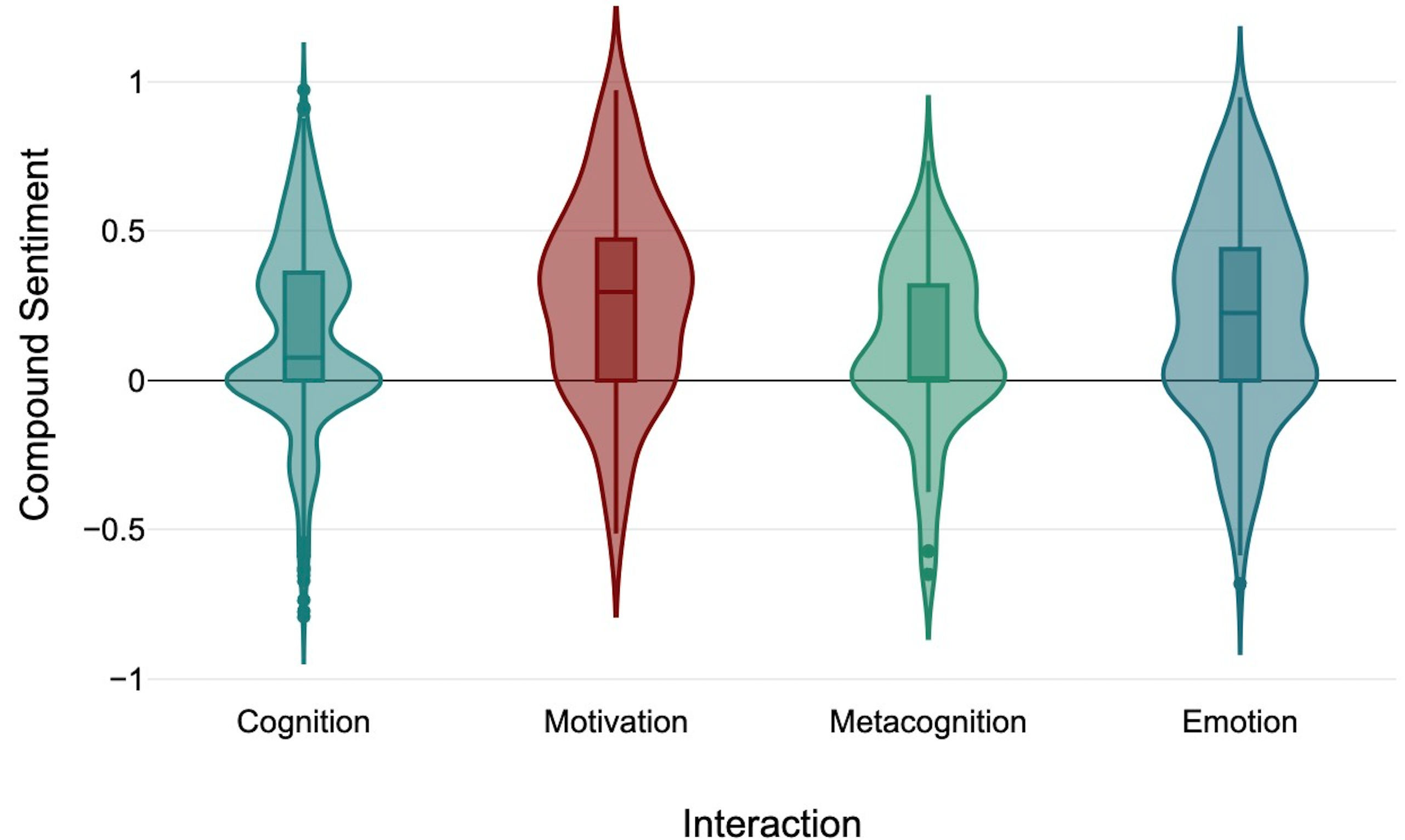} 
        \caption{Compound Sentiment by Interaction}
    \end{subfigure}
    \caption{Interaction by team climate and compound sentiment by interaction.}
    \label{fig:teamwork}
    \vspace{-5mm}
\end{figure}

\vspace{-2mm}
\subsection{RQ2. Triggers to Team Emotional Climate Changes} 
\vspace{-1mm}
We examined differences in emotional climate shifts during decision-making between high-performing (Group A) and low-performing (Group B) teams. In Group A, heart-rate changes were observed as team members navigated complex medical terminology and decision-making related to diagnostic assumptions (\textit{P1} and \textit{P2} are short for \textit{participant 1} and \textit{participant 2} respectively):
\vspace{-1mm}
\begin{tcolorbox}[width=1\textwidth]
(P1): ``Okay. So then, what do you think is going on? \textbf{[searching in the BioWorld library]} ...something in the adrenal glands? Okay, click.'' \textcolor{blue}{\textbf{[heart rate change detected]}}

(P2): ``Catecholamine.''

(P1): ``And then... let's see. So headache, sweating, and tachycardia.''
 
(P2): ``Yeah, pretty exciting. \textbf{[reading on the searching result from the library]}''\textcolor{blue}{\textbf{[heart rate change detected]}}
\end{tcolorbox}
\vspace{-1mm}
This sequence shows that when the team reached a pivotal decision point that they summarized in their report, such as identifying the test results with related symptoms (``elevated catecholamines'' and ``headache, sweating, and tachycardia''), heart rate changes occurred. This suggests engagement during these critical exchanges are necessary for advancing their problem-solving process. Additionally, the ITS assists in knowledge acquisition, helping the team develop a deeper understanding of the case.

Furthermore, heart-rate changes were detected during the following conversation. In this conversation, P1 was seeking clarification about some lab results, specifically values related to measurements which are not fully understood. P2 provided clarification, stating that the values are not elevated, and emphasizes the importance of normal thyroid-stimulating hormone (TSH) levels. The conversation reveals P1’s relief upon understanding that their results are within acceptable ranges and reassures them that there’s nothing concerning beyond normal checks. They also mentioned a key test to order for this given patient profile – TSH, which is an important cue for them to correctly diagnose. 
\vspace{-1mm}
\begin{tcolorbox}[width=1\textwidth]
(P1) ``And then it's... 1560 for the free (urine test). Ah, it said the total was elevated, I don't know. and then 1600 is good?''

(P2) ``These are micrograms. We're not in micrograms. They just aren't elevated.'' \textcolor{blue}{\textbf{[heart rate change detected]}} ...

(P1) ``And then, what helped us?''

(P2) ``The catecholamine [laughs]''

(P1) ``I know.''

(P2) ``I mean, for sure, the fact that everything else is good. Like, for sure, the fact that your TSH is normal is helpful. And then, like the cortisol. This one is also helpful. I think the rest is kinda just, like, make sure they're fine.'' \textcolor{blue}{\textbf{[heart rate change detected]}}

(P1) ``Well, it's good.''
\end{tcolorbox}
\vspace{-1mm}
In contrast, Group B experienced emotional fluctuations during critical knowledge exchange moments but failed to turn them into productive learning events:
\vspace{-1mm}
\begin{tcolorbox}[width=1\textwidth]
(P2) ``Now that everything is thrown back to normal, I think your anxiety one (i.e., the highlighted patient symptom for diagnosis) is higher on this.''

(P1) ``It’s good. Okay. Panic attacks (an incorrect diagnosis was raised).''
\end{tcolorbox}
\vspace{-1mm}
Further discussion about whether other symptoms were relevant to their final hypothesis during which another emotional fluctional was detected:
\vspace{-1mm}
\begin{tcolorbox}[width=1\textwidth]
(P2) ``Like dizzy and faint, not necessarily.''

(P1) ``I agree.  Palpitations, yes.''

(P2) ``Yeah. You're trying to bring it back up?''

(P1) ``Sweating? I am not sure.''
\end{tcolorbox}
\vspace{-1mm}
The dialogue reveals lingering uncertainties and unresolved issues, which led to a misdiagnosis. Their failure to align their emotional shifts with focused problem-solving discussions. Intervention of hints or regulation are therefore needed to prevent negative emotions and foster clarity in decision-making. 

Most of heart rate changes in both groups occurred during critical knowledge exchanges (sharing, refining, or questioning). However, Group A's emotional fluctuations aligned with confident knowledge acquisition, while Group B's emotional fluctuations coincided with uncertain and unsolved inquiries.
\vspace{-2mm}
\section{Discussion and Conclusion}
\vspace{-2mm}
\subsection{Motivational Expressions Drive Positive Team Climate} 
\vspace{-1mm}
Our findings reinforce the critical role of expressions of motivation in teamwork. During team decision-making, the ``mission possible'' mentality—where team members believe in the attainability of their goals—helps sustain motivation and positively influences the team’s emotional well-being. By fostering these socio-motivational interactions, ITSs could be designed to better support positive team experiences, thus leading to enhanced teamwork.
\vspace{-2mm}
\subsection{Emotional Fluctuations as Catalysts in Team Decision-Making} 
\vspace{-1mm}
The results indicate that shifts in emotional climate occur during critical moments of decision-making, where the succeed team took benefits from knowledge sharing and professional skill transfer. These moments are essential for learning, as they highlight the interaction-driven nature of teamwork, where team members actively contribute to each other's learning process. Emotional fluctuations also indicate cognitive engagement and deeper processing of new information. However, only by appropriate regulation and critical reflection within the team, the team potential signaled by the team's emotional shift could be maximized. Without deliberate regulation, emotional shifts risk tipping from productive curiosity into emotions of disengagement (e.g., boredom) or overload (e.g., anxiety). 

For teams that struggled, additional prompts could provide supports to teamwork effectiveness. Providing real-time, adaptive feedback that prompts reflection during moments of emotional highs or lows can guide teams toward more structured, collaborative decision-making. For instance, customized scaffolds that encourage re-evaluation of evidence when emotional tension or uncertainty arises could help mitigate the negative impact of unresolved discrepancies. 

These interventions can help struggling teams to manage emotional variability constructively, turning moments of doubt into opportunities for deeper cognitive engagement and collective problem-solving.

\vspace{-2mm}
\subsection{Implications for ITS Design and Teamwork Research} 
\vspace{-1mm}
These findings underscore the importance of integrating artificial emotional intelligence in ITS. Specifically, future ITS could integrate features that detect and respond to social interaction cues, encouraging interactions that boost team emotional climate. Additionally, design systems to recognize and support periods of emotional fluctuation as part of the learning process, providing scaffolds that guide teams through challenging moments while capitalizing on opportunities for deeper engagement during knowledge exchange. For example, future design can combine emotion detection with behavioral logs to enable affect-aware tutoring, and to provide hints or meta-cognitive prompts such as:
\vspace{-1mm}
\begin{itemize}
    \item \textit{``Take a moment to reflect on and discuss about the differing opinions.''}
    \item \textit{``Do you need to revisit the key information to resolve this uncertainty?''} 
\end{itemize}
\vspace{-1mm}

Furthermore, analyzing how team effectiveness evolves with fluctuations in cognitive and affective states offers valuable insights into team dynamics. Applying this analytical lens to high-stakes teamwork environments offers unique opportunities to: (1) uncover challenge-dependent teamwork mechanisms, (2) identify inflection points where performance pathways diverge, and (3) develop contingency-based strategies for sustaining effectiveness under stress. This approach moves beyond static team assessments toward a process-oriented understanding of adaptive collaboration within technology-rich environments.

\bibliographystyle{splncs04}
\bibliography{references}

\end{document}